**Charge transport mechanism in networks of armchair graphene nanoribbons**

*Nils Richter[1,2], Zongping Chen[3,4], Alexander Tries[1,2,3], Thorsten Prechtl[3,5], Akimitsu Narita[3], Klaus Müllen[2,3,5,†], Kamal Asadi[3], Mischa Bonn[2,3], Mathias Kläui[1,2,\*]*

[1]Johannes Gutenberg-Universität, Institut für Physik, Staudinger Weg 7, 55128 Mainz, Germany
[2]Graduate School Materials Science in Mainz, Staudingerweg 9, 55128 Mainz, Germany
[3]Max Planck Institut für Polymerforschung, Ackermannweg 10, 55128 Mainz, Germany
[4]School of Materials Science and Engineering, Zhejiang University, 38 Zheda Road, 310027 Hangzhou, China
[5]Johannes Gutenberg-Universität, Institut für physikalische Chemie, Duesbergweg 10–14, 55128 Mainz, Germany

†E-mail: muellen@mpip-mainz.mpg.de
*E-mail: klaeui@uni-mainz.de

**Abstract**: In graphene nanoribbons (GNRs), the lateral confinement of charge carriers opens a band gap, the key feature to enable novel graphene-based electronics. Successful synthesis of GNRs has triggered efforts to realize field-effect transistors (FETs) based on single ribbons. Despite great progress, reliable and reproducible fabrication of single-ribbon FETs is still a challenge that impedes applications and the understanding of the charge transport. Here, we present reproducible fabrication of armchair GNR-FETs based on a network of nanoribbons and analyze the charge transport mechanism using nine-atom wide and, in particular, five-atom-wide GNRs with unprecedented conductivity. We show formation of reliable Ohmic contacts and a yield of functional FETs close to unity by lamination of GNRs on the electrodes. Modeling the charge carrier transport in the networks reveals that this process is governed by inter-ribbon hopping mediated by nuclear tunneling, with a hopping length comparable to the physical length of the GNRs. Furthermore, we demonstrate that nuclear tunneling is a general charge transport characteristic of the GNR networks by using two different GNRs. Overcoming the challenge of low-yield single-ribbon transistors by the networks and identifying the corresponding charge transport mechanism puts GNR-based electronics in a new perspective.

Field-effect transistors (FETs) based on two-dimensional materials have attracted immense interest as potential next generation of electronics with exceptional properties, such as mechanical flexibility and optical transparency. Various novel two–dimensional materials have recently become available[1], and among them, graphene plays a unique role due to its extremely high charge carrier mobility[2]. While graphene itself does not have a band gap, a prerequisite for many semiconductor applications, geometrical confinement to one dimension, as occurring in graphene nanoribbons (GNRs), allows for the modification of the electronic structure, and a band gap opening[3–6]. Both the edge structure and the width of the ribbons determine the band gap. The bottom-up synthesis of GNRs provides access to atomically accurate systems, starting from specifically designed molecular precursors, and has therefore attracted much attention[7–11]. The precision of the synthesis together with the versatility of different GNR structures make them ideal objects for testing theoretical predictions concerning their electronic and magnetic properties[3,5,12]. Especially the use of the chemical vapor deposition (CVD) method allows for the scalable, cost-effective and high-throughput production of high-quality GNR films[13,14]. Among those GNRs achievable by CVD[7], ribbons with armchair edges and a width of five carbon atoms[14,15] (5-AGNRs) are particularly interesting for charge transport because they exhibit a particularly low band gap, as predicted theoretically and confirmed spectroscopically[4–6,16]. Furthermore, photoconductivity measurements on 5-AGNRs indicate very high mobility of charge carriers in these nanoribbons[15,17]. Despite the evident promise of 5-AGNRs, they have not been well studied in devices[14]. And, most importantly, many of the fundamental charge transport mechanisms of GNRs have so far remained unclear. Up to now, most device studies on bottom-up GNRs have aimed at observing charge transport through single ribbons, employing short-channel FETs[18–20]. While these devices show promise for nano-electronics applications, they are typically highly resistive, which has been attributed to large energy barriers at the contacts for charge injection[13,18,20–22]. The recent use of nine-atom-wide AGNRs (9-AGNRs) has enabled substantial improvement regarding conductivity and electrostatic current modulation compared to other previously used GNRs[20]. Still, the challenge of fabricating reliable and reproducible devices based on isolated GNRs, impedes the device characterization such as the extraction of Schottky-barrier heights at the contacts. As alternative to short-channel devices, networks or arrays of GNRs can be used as a basis for a field-effect transistor[23,24]. Such thin film devices have several advantages, including a scalable fabrication process, providing a broad scope of applications, not only as transistors but also, for example, in optoelectronic and chemical sensing. Here, we show that GNR network-based devices provide a means to elucidate the physics underlying inter- and intra-ribbon charge transport.

We investigate charge transport in networks of bottom-up synthesized 5-AGNRs and 9-AGNRs grown by CVD. The 5-AGNR serves as a model system to identify the relevant charge transport mechanism in networks, as they are highly conductive with a resistance two orders of magnitude lower than that of 9-AGNR. We have developed a dependable and reproducible device fabrication protocol, which enables reliable transport measurements of the GNR networks and allows a direct comparison of the two ribbon types. In contrast to single-ribbon devices with Schottky-barrier dominated contacts[13,18,20], the charge injection is not limited in our network FETs with Ohmic contacts, allowing for detailed electrical characterization. Measurements over a previously inaccessible wide temperature range reveal the dominant charge transport mechanism to be nuclear tunneling-assisted carrier hopping. Based on this model, the universal scaling allows us to collapse all the charge current characteristics obtained across several orders of magnitude of bias voltages and temperatures onto a single curve. From this curve, we determine a consistent charge carrier hopping distance.

**Results**

**Device fabrication. Figures 1 (a)** and **(b)** present a schematic depiction and an optical micrograph of a typical GNR network FET device. A heavily doped silicon wafer served simultaneously as substrate and back gate electrode. The gate electrode was separated from the lateral GNR channel by a 300-nm-thick silicon oxide layer. For device fabrication, electron beam lithography was used to define 25-nm-thick Au source and drain electrodes with a thin Cr layer (5 nm) as an adhesion layer. The channel length was varied from approximately 500 nm to 5 μm, and the channel width was set to a constant value of 500 μm. Finally, the CVD-grown 5-AGNR films and 9-AGNR films, respectively, corresponding to a monolayer of ribbons, were transferred on top of the electrodes. This technique has been used already in the past successfully for upscaling graphene transistors[25]. To synthesize and transfer the films, our previously reported[13,15] techniques were employed (for more details on the device fabrication see also the Methods section and Ref. 26). The ribbons within the films lie close to one another, forming a densely packed ribbon network as revealed by scanning tunneling microscopy (STM) of the films as-prepared on the Au growth-substrate[15,27]. Hence, the topography allows for inter-GNR charge carrier transfer, and therefore macroscopic charge currents can be established via percolation paths. Raman spectroscopy demonstrates the integrity of the GNR network transferred on top of the device structures, since the Raman spectra display identical peaks before and after the transfer (**Fig. 1 (c)** for 5-AGNRs and **Fig. S1** in the **supplemental information**[28] for 9-AGNRs). Moreover, the width of the GNRs constituting the films can be unambiguously confirmed by the Raman response associated with the radial breathing-like mode (RBLM)[15,29].

The intense RBLM peak at 533 cm$^{-1}$ can be assigned to 5-AGNRs. Additionally, a small RBLM peak is visible at 283 cm$^{-1}$, indicating the presence of 10-AGNRs with double the width of 5-AGNRs. A small fraction of 10-AGNRs is formed by a lateral fusion of two 5-AGNRs at the annealing temperature of 400 °C, as has been previously reported by us[15]. One can, however, neglect the contribution of the small number of 10-AGNRs to the charge transport experiments as they have a larger band gap than 5-AGNRs[15].

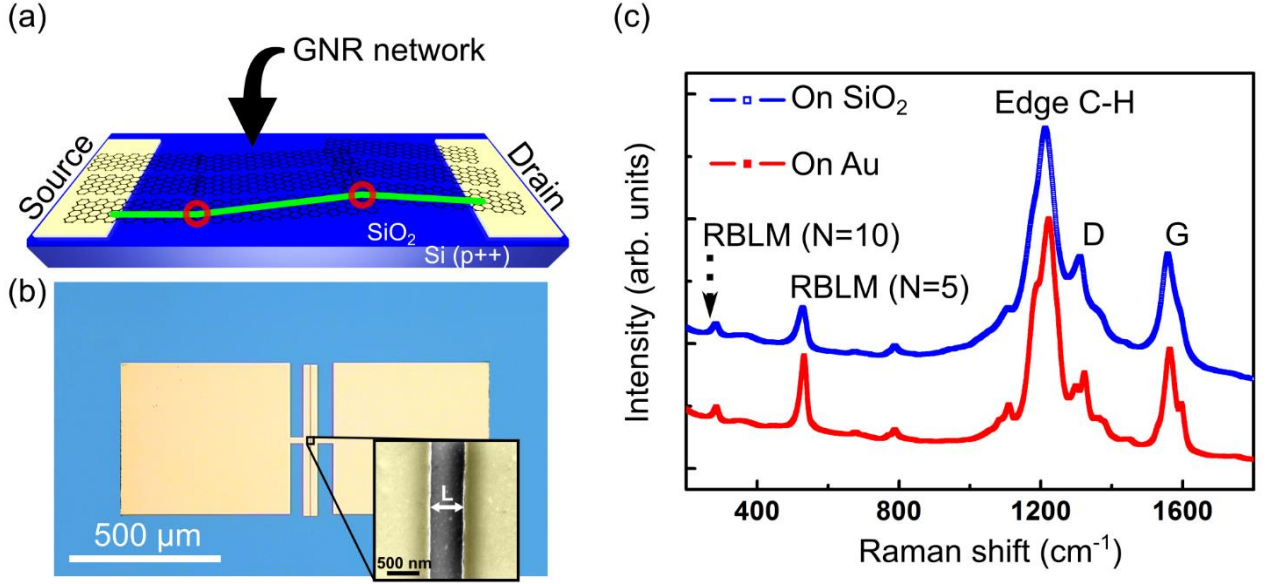

**Figure 1**: (a) Schematic depiction of a GNR network FET where a possible percolation path is drawn in green. The red circles mark locations of charge transfer between the densely packed ribbons. The current flows between the metallic (Au) source and drain electrodes through the GNR channel. The GNR film covers the whole substrate surface. The drawing is not to scale. (b) Optical micrograph of a GNR network FET. The SiO$_2$ surface appears blue, while the metallic contacts are golden. The inset shows scanning electron microscopy image magnifying the channel region where charge current flows through GNR networks (Au electrodes false-colored in yellow). The junction has a separation of $L \approx 600$ nm. (c) Raman spectrum of a 5-AGNR film before and after the transfer from an Au substrate to a SiO$_2$ surface. The spectrum exhibits the usual D (at approximately 1340 cm$^{-1}$) and G (between 1565 cm$^{-1}$ and 1595 cm$^{-1}$) peaks of crystalline sp$^2$ carbons. The peak at approximately 1220 cm$^{-1}$ indicates the presence of carbon-hydrogen bonds, located along the periphery of all ribbons. The low-frequency lines can be attributed to the width-dependent RBLMs[15], where the width is denoted by $N$, the number of carbon atoms across the ribbon.

**Electrical characterization of the network**. At room temperature and in an inert gas atmosphere, we measured the FET output characteristics of the 5-AGNR network by fixing $V_G$, the voltage applied to the gate electrode, and sweeping the drain voltage $V_D$ from 100 mV to 20 V and back, with grounded source electrode. The drain current $I_D$ was measured, with parasitic leakage currents through the gate oxide negligible compared to $I_D$ as shown in **Fig. S2**. To avoid artifacts, we ensured the drain current to be always much larger than the gate leakage current. Representative output curves at five different gate voltages from $+70$ V to $-70$ V are presented in **Fig. 2 (a)**. Clearly, different gate biases modulate the conductance of the network. The current increases linearly with increasing drain voltage for $V_D < 1$ V with a reciprocal slope $R_{on}$ and follows a power law $I_D \propto V_D^\beta, \beta > 1$ for larger voltages (see further modelling below). Hysteresis in the output curves was negligible for all measurements. The transfer characteristics of the GNR-network FET are shown in **Fig. 2 (b)**. The device showed minor hysteresis between for- and backward sweeps. The current modulation ratio $I_{VG,1/VG,2}$ and the field-effect mobility $\mu_{FE}$ calculated from the transfer curves (see **S3** for more details on these definitions) amounted to $I_{-80\,V/+80\,V} \approx 5$ and $\mu_{FE} \approx 2 \times 10^{-2} \mathrm{cm^2 V^{-1} s^{-1}}$, respectively. The current modulation ratio of the 5-AGNR network is low due to the presence of a high background of charge carriers. Using Ohm's law, this background density can be directly estimated $n_0 = eI_D\mu_{FE}/V_D = 2 \times 10^{12}\,\mathrm{cm^{-2}}$ (at $T = 260$ K), which is comparable to the electrostatically induced charge carrier density $n_{\mathrm{ind}} = C_{\mathrm{Ox}}V_G/e \approx 3.6 \times 10^{12}\,\mathrm{cm^{-2}}$ (at $V_G = 50$ V). Since the expected band gap of ~1.7 eV[5] is much larger than the thermal energy $k_BT$ at room temperature, the large charge carrier density can be attributed to extrinsic doping[30].

The transfer characteristics of the 9-AGNR network FETs are shown in the inset of **Fig. 2 (b)**. For the shown device, we find $\mu_{FE} \approx 1 \times 10^{-3} \mathrm{cm^2 V^{-1} s^{-1}}$, and a current modulation of $I_{-60\,V/+60\,V} \approx 120$, the latter is comparable to the reported short-channel devices of 9-AGNR with thick SiO$_2$ gate barriers[20].

Formation of low resistance Ohmic contacts between the GNR network and the FET electrodes is crucial for the reliable extraction of the transport properties[31]. Hence, we analyzed the channel length scaling of the output and transfer curves to retrieve information on the Au/5-AGNR junction. To this end, the transmission line method (TLM)[32] was applied. A typical TLM plot is shown in **Fig. 2 (c)** for three different gate voltages (−60 V, 0 V and +60 V. For 5-AGNR FETs, the normalized contact resistance $R_CW$ varies between ~210 Ωm at $V_G = -60$ V and ~320 Ωm for $V_G = 0$ and higher. We note, that $R_C$ showed a slight dependence on the gate bias as shown in **S4** of the **supplemental information** similar to carbon nanotubes[33]. Furthermore, at zero gate voltage, the ratio of the contact resistance versus the total channel resistance (for $L = 1\ \mu m$) is $< 0.2$. Therefore, the channel conductance is dominated by the conductance of the 5-AGNR network and not by the sporadic contact resistance, which encourages the following analyses.

The FET fabrication process is robust with a high device yield approaching 100 %. This is evidenced by a comparison of tens of devices on different chips for both types of GNRs (in total 43 devices for 5-AGNRs and 20 devices for 9-AGNRs). **Figure 2 (d)** displays histograms of the relative frequency for three important transport quantities for all 5-AGNR devices, which have the same channel length: The exponent $\beta$, the Ohmic resistance $R_{on}$ and the mobility $\mu_{FE}$. The narrow distribution for each of these device parameters highlights the high homogeneity and uniformity of the GNR networks, and device reproducibility. Especially, the low error for the mean value of $\beta$ of only 2 % is noteworthy, since $\beta$ is directly connected with the underlying charge transport mechanism (as detailed below).

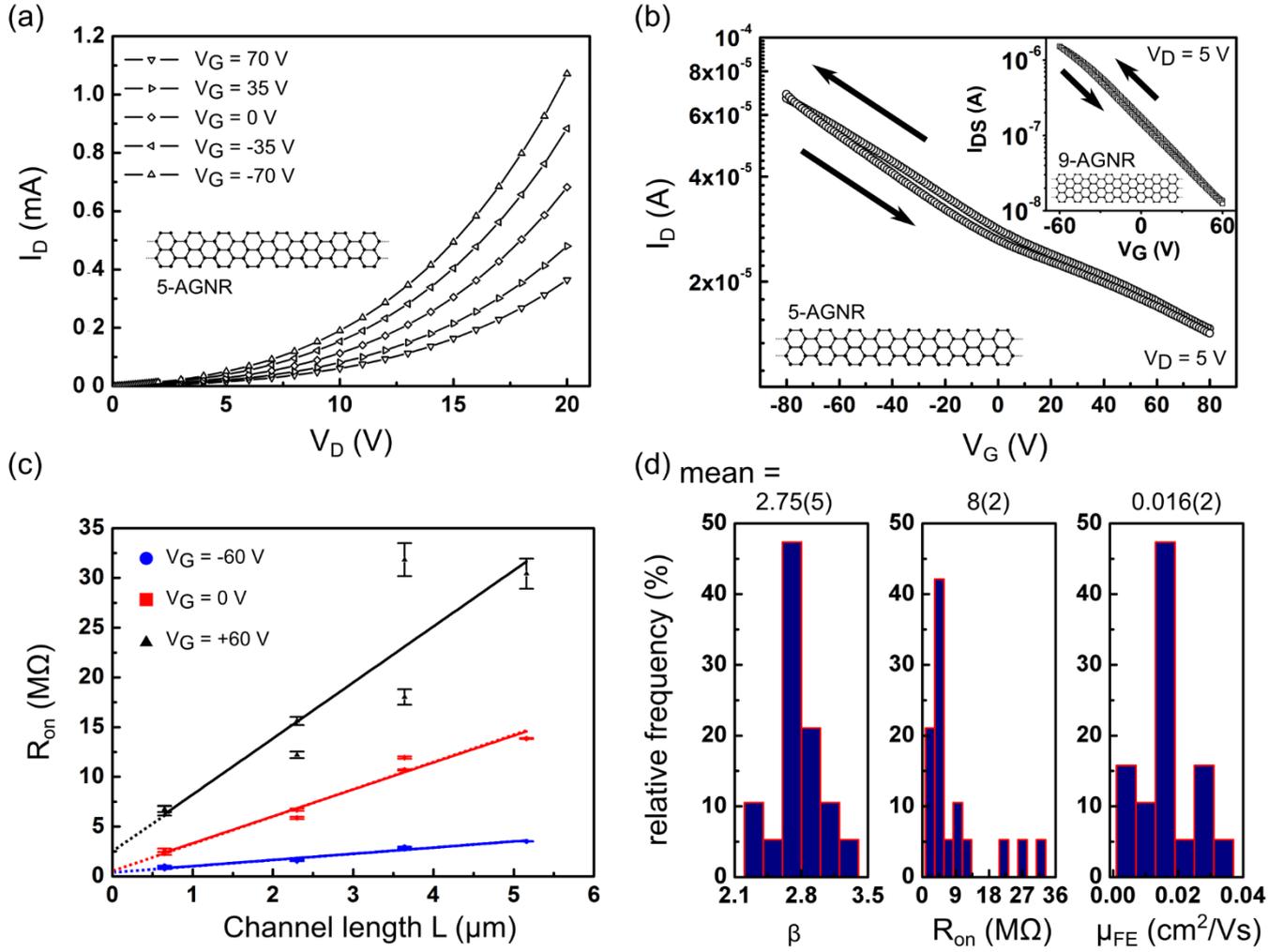

**Figure 2**: Plot of output (a) and transfer curves (b) of 5-AGNR. The channel current $I_D$ responds in an Ohmic–like fashion at low $V_D$ ($V_D \leq 1$ V). In (b), the arrows indicate the sweep direction and the inset shows a transfer curve measured for a 9-AGNR network device, which is roughly two orders of magnitude less conductive. Lines in (a) are guides for the eye. (c) Total device resistance (channel and contact resistance) $R_{on}$ as a function of channel length for contact resistance extraction at different gate voltages. Solid lines are linear fits to the data and the dotted lines show the extrapolation to zero channel length, indicating contact resistance. (d) Device parameter spread at room temperature. Relative frequency of the values for the exponent $\beta$, the Ohmic resistance $R_{on}$ and the mobility, measured for 19 5-AGNR devices with the same channel length. Mean values and statistical errors of the last digit of the mean values (in parentheses) are indicated above the histograms.

**Universal scaling of charge transport in graphene nanoribbon networks.** Having established the reproducibility of 5-AGNR network devices, we proceed to investigate the charge transport mechanism. The precise mechanism of charge transport is not a priori evident since the channel length is much longer than the length of individual nanoribbons. Hence, charge carriers must cross ribbon–ribbon junctions in the GNR networks to allow for macroscopic current, suggesting that inter-GNR hopping will contribute to the overall transport. The temperature dependence of charge transport can help to identify the nature of the charge transport mechanism in the network. Therefore, a helium bath cryostat was used to measure the output characteristics from ~262 K down to ~5 K ($V_G = 0$ V) for a device with channel length $L = 1$ μm (**Fig. 3 (a)**). At low voltages, the transport is Ohmic, while at higher voltages, the current grows superlinearly with $V_D$, following a power law $I_D \propto V_D^\beta$, with $\beta = 2.76 \pm 0.04$ at $T = 262$ K and zero gate voltage, where gate modulation leaves these functional dependencies unchanged. Such behavior indicates that the charge transport mechanism is through inter-ribbon hopping[34,35]. To describe the charge transport, we employ a quantum mechanical model of dissipative tunneling in a biased double well, mediated by nuclear vibrations, which act as a heat bath[36,37]. In this so-called nuclear tunneling mechanism, the coupling of the electronic charges to their nuclear environment defines the potential energy landscape for charge motion. Intuitively, in the low bias regime, charge transport occurs predominantly by tunneling through the energy barrier, and it is temperature dependent because of the coupling of the charge to the nuclear environment. At sufficiently high bias however, the double well becomes so asymmetric that the charge carrier can overcome the energy barrier at no energy costs, and therefore the transport becomes virtually temperature independent.

The hopping rate equation and the resulting current have been derived by Fisher and Dorsey, and Grabert and Weiss[36,37], to read

$$I_D = I_0 T^{1+\alpha} \sinh\left(\frac{\gamma eV}{2k_B T}\right) \left| \Gamma\left(1 + \frac{\alpha}{2} + i\frac{\gamma eV}{2\pi k_B T}\right) \right|^2, \tag{1}$$

where $\gamma^{-1}$ is the number of hops that a charge experiences when travelling from one electrode to the other, $\alpha$ is a scaled version of Kondo parameter describing the coupling between the charges and the heat bath, and $\Gamma$ represents the complex gamma function. In the limit $\lim_{V \to 0}$, the current is Ohmic and given by:

$$\lim_{V \to 0} I_D = \frac{I_0 \gamma e}{2 k_B T} \left| \Gamma \left( 1 + \frac{\alpha}{2} \right) \right|^2 T^\alpha V, \tag{2}$$

For $\lim_{V \to \infty}$, the current is temperature independent, and has a power law dependence on voltage:

$$\lim_{V \to \infty} I_D = I_0 \pi^{-\alpha} \left( \frac{\gamma e}{2 k_B} \right)^\beta V^\beta, \tag{3}$$

The exponents $\alpha$ and $\beta$ have to follow the relation $\beta = \alpha + 1$. We determined the exponent $\alpha$ from the temperature dependent linear part of the curves using **Eq. 2**: $\text{Log}(J_D)$ as a function of $\log(T)$ at different drain voltages is plotted in **Fig. 3 (b)**, where $J_D = I_D(L/W)$. The slopes of linear models for each $V_D$ are obtained by a least squares fit giving in the limit of vanishing drain voltage $\lim_{V \to 0} \alpha = 3.0 \pm 0.1$. At high bias and low temperature, we experimentally find that the exponent $\beta$ in **Eq. (3)** becomes $\beta \approx 4$, validating the prediction $\beta = \alpha + 1$ and corroborating the applicability of this model. Together, these experimental findings suggest that all measurements of $I_D$ at different temperatures and voltages can be combined to a single curve, when $J_D/T^{\alpha+1}$ is plotted as a function of relative energy $eV/k_B T$ as demonstrated in **Fig 3 (c)**. This plot consists of 1552 data points which stem from the output curves between 262 K and 46 K and from a temperature sweep at a fixed high bias voltage $V_D = 10$ V down to 5.6 K (see **Fig. S5**). By fitting **Eq. 1** to the scaled curve with $J_0$ and $\gamma$ as the only free fitting parameters, we found excellent agreement with $J_0 = (8.3 \pm 0.2) \times 10^{-13}$ A K$^{-\alpha-1}$m$^{-1}$ and $\gamma = (17.6 \pm 0.2) \times 10^{-3}$, over the entire range of gate voltages in our experiment.

To check the universal character of this analysis, the temperature dependent charge transport experiments were repeated for an FET based on 9-AGNR networks with a channel length of $L = 2$ μm. Here as well, the universal scaling is applicable with $\alpha = 9$, $J_0 = (2.1 \pm$ 0.1) × $10^{-32}$ A K$^{-\alpha\text{-}1}$m$^{-1}$ and $\gamma = (9.7 \pm 0.1) \times 10^{-3}$. Finally, the dependence of the hopping rate $\gamma$ can serve as an independent check of our conclusion, that inter-GNR hopping limits charge mobility in our network devices. The product $L \times \gamma$ yields the statistical average of a hopping length, which for both, the 5-AGNR and the 9-AGNR points to a distance between charge carrier hops of 17 to 19 nm. Interestingly, this hopping length is comparable to the average length of individual GNRs in the network[13,15]. This geometrical agreement implies that the limiting factor for charge transport in the network is inter-ribbon hopping. The intra-ribbon mobility on the other hand can be orders of magnitude larger, supported by observed band-like transport in the GNRs[8,13,15–17,34,38,39].

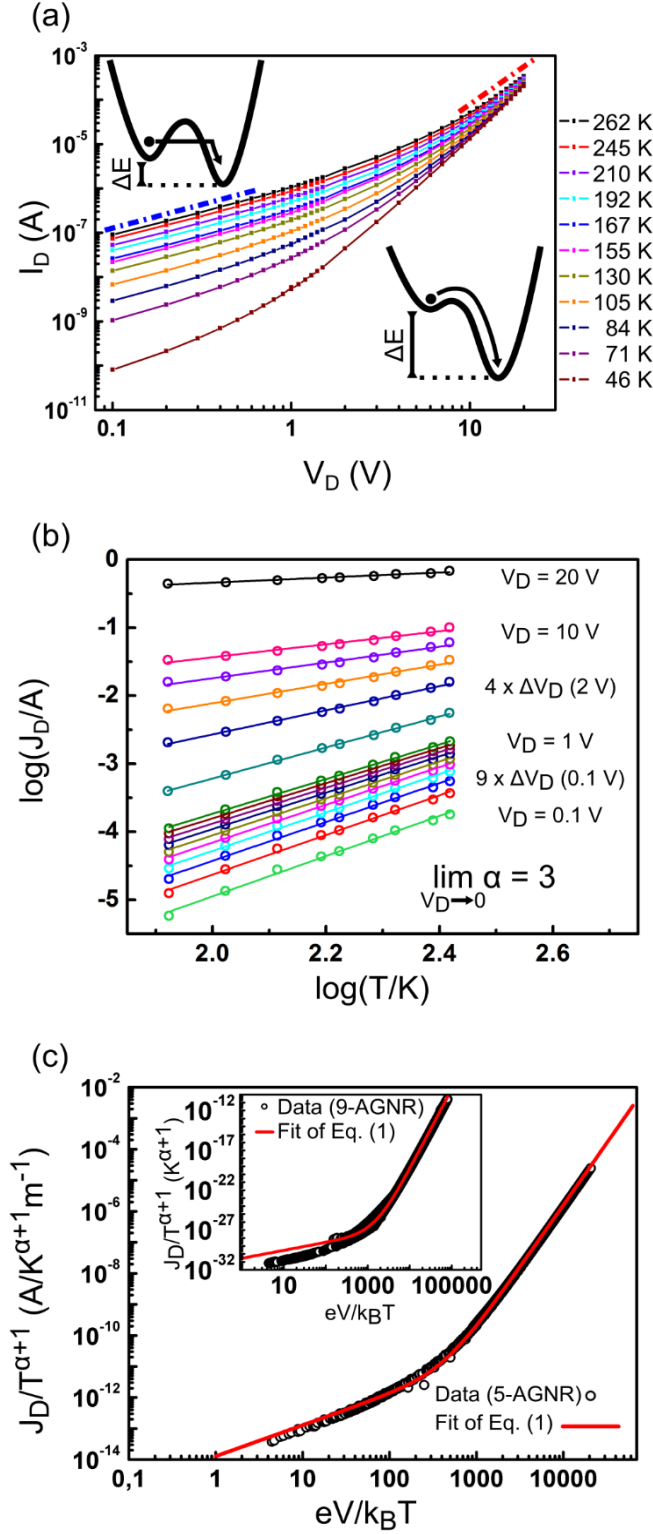

**Figure 3**: Temperature dependence of charge transport in 5-AGNR networks. (a) shows the evolution of output curves with temperature. Solid lines are guides for the eye. The dashed lines indicate the linear low bias regime (blue) and non-linear high bias regime (red). The charge transport mechanism for low and high bias is shown schematically. In (b), we show cuts through the output curves for fixed drain voltages as indicated in the figure. Here, lines represent linear fits through the data to determine the exponent of a power law $I_D \propto T^\alpha$. In (c) we plot the scaled channel current density $J_D/T^{\alpha+1}$ as a function of relative energy $eV/k_B T$. The solid red line is a fit of **Eq. 1** with excellent agreement with the measurement. The inset shows the universal scaling curve for 9-AGNRs.

**Conclusion**

We have demonstrated reproducible FETs based on networks of 5- and 9-AGNRs. The network-based FETs do not rely on identifying and contacting individual ribbons and are therefore able to boost the fabrication yield. The device properties are reproducible, with a narrow spread in parameters. The FETs based on 5-AGNR showed unprecedented conductance two orders of magnitude larger than that of 9-AGNR. Through a systematic analysis of $I(V,T)$ characteristics, we were able to determine the nature of charge transport characteristics in both GNR networks. We found that the transport characteristics can be described by a universal scaling based on a fully quantum mechanical hopping transport mechanism. For both 5- and 9-AGNRs, the different $I(V,T)$ characteristics could be collapsed onto a single universal curve, indicating the generality of the transport mechanism. The universal curves integrated measurements at temperatures between 5 K to 262 K for voltages, swept over two orders of magnitude. The modeling determined the hopping of the charge carriers between nanoribbons as the factor limiting charge transport. The use of long GNRs in networks will therefore enable higher field-effect mobilities for future GNR-based FETs.

**Methods**

**Growth of graphene nanoribbons by chemical vapor deposition**. All 5-AGNRs and 9-AGNRs were synthesized from surface-assisted chemical vapor deposition (CVD) technique as we reported previously elsewhere[13,15,27]. The CVD system comprises a horizontal tube furnace (Nabertherm, RT 80-250/11S) and heating belt (Thermocoax Isopad S20). The Au/mica substrate was loaded into the tube furnace as the growth substrates and heated to 250 °C under a gas flow of Ar (500 sccm) and $H_2$ (100 sccm) with a pressure of ~1.5 mbar. The precursor for 5-AGNRs, an isomeric mixture of 3,9-dibromoperylene and 3,10-dibromoperylene (DBP), was then sublimed by the heating belt at ~250 °C and deposited on the Au/mica substrate for 30 min for polymerization and subsequently annealed at 400 °C for 15 min for cyclodehydrogenation. Similarly, for the synthesis of 9-AGNRs, the Au/mica substrates were loaded into the tube furnace and heated to 200 °C under a gas flow of Ar (500 sccm) and $H_2$ (100 sccm) with a pressure of ~1.5 mbar. At the meantime the monomer 3',6'-dibromo-1,1':2',1''-terphenyl was loaded upstream and sublimed at 150 °C for 30 min for polymerization. Subsequently, the samples were annealed at 400 °C for 15 min for cyclodehydrogenation.

**Device fabrication**. The heavily doped silicon wafers, which served as both substrate and back gate electrode, are commercially available and have a 300 nm thick silicon oxide layer (thermally oxidized). The wafers have been diced into chips of $1 \times 1$ cm$^2$. We used electron beam lithography to define 25 nm thick Au source and drain electrodes where a thin layer Cr (5 nm) served as an adhesion layer. Finally, GNR films were transferred on top of the structures. In the case of 9-AGNRs, we transferred two layers.

**Transfer of GNR thin films.** The procedure to transfer the films, we employed a technique which we have reported previously[13]: After the CVD growth of the GNRs, a thin layer of poly(methyl methacrylate) (PMMA) was spun onto the GNR/gold/mica stack which provided additional mechanical stability and facilitated the transfer of intact films over a large area. Carefully, the resulting stack was floated on concentrated HF for several hours to delaminate the PMMA/GNR/gold film from the mica slab. After the delamination was complete, the gold was etched away in a gold etchant (Sigma-Aldrich). We then transferred the PMMA/GNR film to the target substrate with Au electrodes. To dissolve the PMMA, the PMMA/GNR/substrate stack was immersed in an acetone bath. Finally, we rinsed the chip with isopropanol and dry blow the GNR film.

**Raman spectroscopy.** Raman characterization of the GNR films was performed with a Bruker SENTERRA RFS100/S Raman spectrometer using a 785 nm laser under ambient conditions.

**Room temperature electrical characterization**. Measurements have been performed using a three-terminal probe station integrated into an inert gas glove box with a nitrogen atmosphere. The probe station was connected to a Keithley 4200 semiconductor characterization system, which contains three independent source-measure units and allows for the electrical characterization of test devices.

**Variable temperature electrical characterization**. Variable temperature measurements have been carried out in a bath cryostat equipped with a dynamic variable temperature insert (VTI) at the bottom of the sample space, which allows for the control of the sample temperature. Stable temperatures within approximately 0.1 K are obtained by balancing the cooling power of a liquid helium flow against the heating power of an electrically resistive heating element. The liquid helium is drawn from the main reservoir through a needle valve, which is adjusted manually. Temperatures below 4.2 K are obtained by reducing the vapor pressure of the liquid helium in the sample space by mechanical pumping. Electrical measurements are enabled with a Keithley 238 source-measure combined with a Keithley 2400 source-measure unit. To accurately measure high resistive samples in this setup, we employed a current guarding method.

**Author contributions:** N. R., A. T. fabricated and measured the devices (under supervision by M. K., M. B. and K. A.). T. P. synthesized the molecular precursors, and Z. C. fabricated and transferred the graphene nanoribbons (under supervision by A. N. and K. M.). Z. C. performed Raman spectroscopy on the GNR films. All authors discussed the results and the paper.

**Acknowledgements:** This work was financially supported by the DFG primarily through the Priority Program Graphene SPP 1459 and partly by SFB TRR 173 Spin+X, the Max Planck Society, the Seventh Framework Programme within the project Moquas Molecular Quantum Spintronics FET-ICT-2013-10 610449 and the US Office of Naval Research BRC Program. We thank Tim Dumslaff for the preparation of the monomer precursor of 9-AGNR. We thank Wojciech Pisula for providing the electrical characterization setup in a nitrogen gas glove box. We thank Paul Blom and Marie-Luise Braatz for stimulating discussions. N. R. gratefully acknowledges the MAINZ Graduate School of Excellence (DFG GSC/266) as well as the Carl

Zeiss Stiftung. A.T. is a recipient of a DFG-fellowship through the Excellence Initiative by the Graduate School Materials Science in Mainz (DFG GSC/266). K.A. is grateful to the Alexander von Humboldt Foundation for the funding provided in the framework of the Sofja Kovalevskaja Award, endowed by the Federal Ministry of Education and Research, Germany.

# Supplemental Information

**Charge transport mechanism in networks of armchair graphene nanoribbons**

*Nils Richter[1,2], Zongping Chen[3,4], Alexander Tries[1,2,3], Thorsten Prechtl[3,5], Akimitsu Narita[3], Klaus Müllen[2,3,5,†], Kamal Asadi[3], Mischa Bonn[2,3], Mathias Kläui[1,2,\*]*

[1]Johannes Gutenberg-Universität, Institut für Physik, Staudinger Weg 7, 55128 Mainz, Germany
[2]Graduate School Materials Science in Mainz, Staudingerweg 9, 55128 Mainz, Germany
[3]Max Planck Institut für Polymerforschung, Ackermannweg 10, 55128 Mainz, Germany
[4]School of Materials Science and Engineering, Zhejiang University, 38 Zheda Road, 310027 Hangzhou, China
[5]Johannes Gutenberg-Universität, Institut für physikalische Chemie, Duesbergweg 10–14, 55128 Mainz, Germany

†E-mail: muellen@mpip-mainz.mpg.de
\*E-mail: klaeui@uni-mainz.de

## S1. Raman spectroscopy of 9-AGNR films

In **Fig. S1** we show Raman spectra taken on a 9-AGNR film as-grown on Au and after the transfer on SiO$_2$ demonstrating the intactness of the film after the transfer. We find the RBLM peak at approximately 311 cm$^{-1}$, which is the expected value for 9-atom wide GNRs[1].

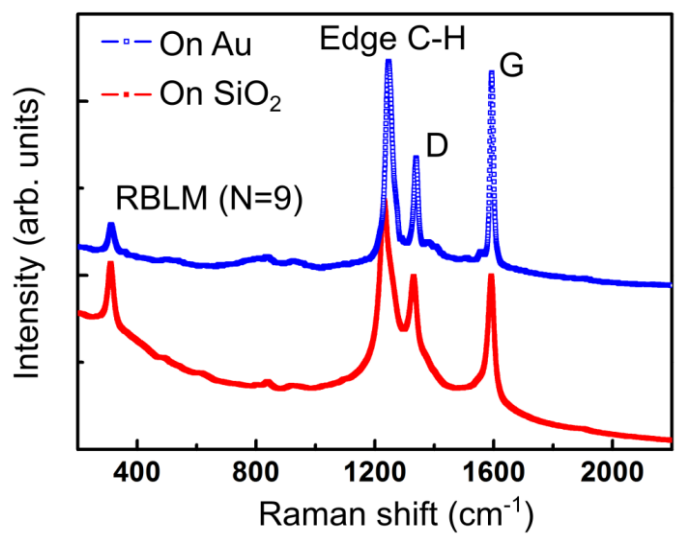

**Figure S1**. Raman spectrum of a 9-AGNR film on Au and SiO$_2$ surfaces. The low-frequency line at 311 cm$^{-1}$ can be attributed to the presence of intact 9-AGNRs (the ribbon width is denoted by $N$, the number of carbon atoms across the ribbon).

## S2. Gate-leakage and bias symmetry in 5-AGNR network FETs

As described in the main text, in FET devices parasitic leakage currents can occur through the gate barrier. At high gate voltages and low drain voltages, the leakage leads to small systematic errors in the measurement. In order to quantify the impact of leakage currents, we measure the current at the gate electrode $I_G = I_{DG} + I_{SG}$, where $I_{SG}$ is the current flowing between the source electrode and the gate electrode. As shown in **Fig. S2**, the gate current is small compared to the drain current and its dependence on $V_D$ is negligible at small voltages applied to the gate electrode. Only at large $V_G$, the leakage starts to affect $I_D$ slightly. However, since $I_G$ does not change with $V_D$, we correct for its influence by simply subtracting $I_G$ from the I-V curves, such that $\lim_{V_D \to 0} I_D = 0$.

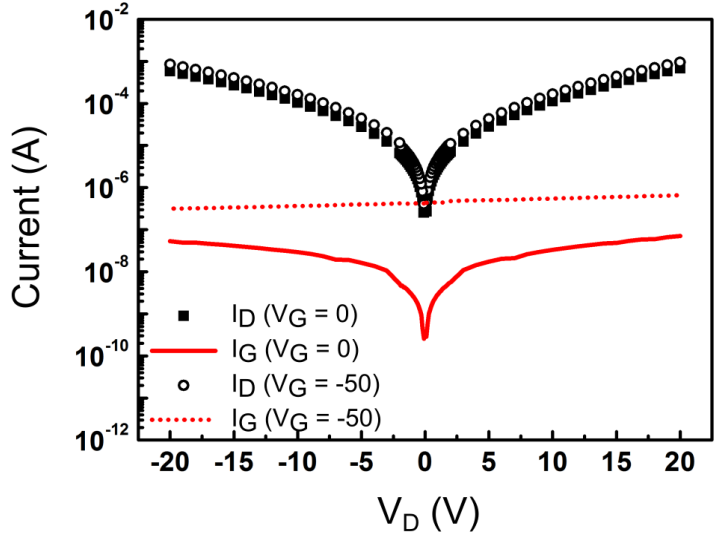

**Figure S2**. Gate leakage current and bias symmetric channel current for a representative 5-AGNR device presented in a semi-logarithmic plot. The gate current is orders of magnitude below the channel current and does not depend on bias voltage. We plot the absolute value of the channel current in order to provide a facile comparison of positive and negative bias. Although the current at negative drain voltages is systematically lower than the current at positive drain voltages, this difference is very small. Hence, the use of only positive bias data is justified.

## S3. GNR-network FET characterization

In a conventional FET, the ratio $I_{\text{on/off}}$, which compares the current in the saturation regime (on-state) with the current in the subthreshold regime (off-state), is widely used as a figure of merit to quantify the effect of the gate voltage. Since a current saturation regime is not reached in our devices, we define in analogy to $I_{\text{on/off}}$ a current modulation ratio

$$I_{VG,1/VG,2} = \frac{I(V_{G,1})}{I(V_{G,2})}, \tag{S1}$$

where we compute the ratio of channel current at specific gate voltages. Furthermore, the field-effect mobility can be extracted from the transfer curve via[2]

$$\mu_{FE} = \frac{g\,L}{V_D\,C_{0x}\,W}, \tag{S2}$$

where $g = \partial I_{SD}/\partial V_G$ is the transconductance, $L$ is the channel length, $W$ is the channel width and $C_{0x} = \epsilon\epsilon_0/t_{0x} \approx 1.15 \times 10^{-4}\,\text{F/m}^2$ is the geometrical capacitance density assuming that the channel and the gate electrode form a parallel plate capacitor. Here, we use $t_{0x} = 300$ nm, a relative permittivity of SiO$_2$ $\epsilon = 3.9$ and a vacuum permittivity $\epsilon_0 = 8.854$ F/m. Ideally, the transconductance is determined in the linear regime of the transfer curve. However, with our GNR network FETs, a linear regime is not reached. Therefore, we use a linear approximation of the curve in the range $V_G \leq -35$ V. In the linear regime of transfer curves, the channel current is usually much larger than in the subthreshold regime and therefore the transconductance, which we extract in this way leads to a systematic underestimation of $\mu_{FE}$. Hence, the values given for $\mu_{FE}$ are lower bounds. However, the contact resistance-free mobilities are of the same magnitude showing that the systematic error in the field-effect mobility is small and hence these values represent a good approximation of the charge carrier mobility in the devices.

With nuclear tunneling-assisted hopping as the dominant charge transport mechanism, we can further rationalize the values for the field-effect mobility. Although a degradation of the charge carrier mobility with larger band gaps is expected, theory predicts values for the mobility in the order of $10^2$ cm$^2$V$^{-1}$s$^{-1}$ for GNRs with a band gap of comparable size[3], much larger than the measured values. Using terahertz spectroscopy, comparable values have been experimentally observed in 9-AGNR samples[1]. On the other hand, when hopping is the dominant charge transport mechanism, mobilities in the range of $10^{-1}$ cm$^2$V$^{-1}$s$^{-1}$ to $10^{-4}$ cm$^2$V$^{-1}$s$^{-1}$ are typically observed[4].

## S4. Gate voltage-dependence of the contact resistance

The gate voltage dependent contact resistances were determined for a series of 5-AGNR devices by measuring charge transport at room temperature for various channel lengths and using the transmission line method[5]. The total device resistance consists of the channel resistance and the contact resistance, $R_{on} = R_{channel} + R_C$. The channel resistance is a function of gate voltage and is proportional to the geometrical aspect ratio of the channel $L/W$, with $L$ the channel length and $W$, the channel width. We use the Ohmic part of the I-V curves at low drain voltages to determine the device resistance $R_{on}$ as the slope of a linear model via a least squares fit. The data points for each gate voltage were again linearly fitted (least squares fit) to extract the total contact resistance (source + drain) $R_C$ from the intercept of the fit curve with the ordinate-axis at zero channel length. In **Fig. 2 (d)** of the main text, we show the result of this procedure for one device at three different gate voltages. The contact resistance $R_C$ is expected to be a constant with respect to the gate voltage. However, due to current crowding effects a gate voltage dependence can be induced[6] and we observe a decrease of $R_C$ when the gate voltage is lowered from large positive values towards zero. When the gate voltage is further lowered towards large negative values, we gradually turn on the conducting channel and the contact resistance becomes constant. This behavior is typical for example in CNT network field-effect transistors[7]. At zero gate voltage, the ratio of contact resistance versus total device resistance (for $L = 1\ \mu m$) is $< 0.2$. Therefore, we conclude that the contact resistance does not play a dominant role our measurements and we can neglect it in our analysis.

To corroborate this further, we exemplify a correction for the charge carrier mobility and compare the corrected value to the as-measured data. With the help of the width–normalized inverse channel resistance $m = (\partial R/\partial L)^{-1}/W$ at different gate voltages (**Fig. S3**) the contact resistance-corrected charge carrier mobility[5] $\mu_{RC} = 1/C_{Ox}(\partial m/\partial V_G)^{-1} = (3.4 \pm 0.0.2) \times 10^{-3}$ cm$^2$V$^{-1}$s$^{-1}$ can be determined. Following this method, any apparent channel-length dependence of the field-effect mobility originating from contact resistance at the source and drain electrodes is eliminated. However, for this set of devices $\mu_{FE}$ extracted from the transfer curves ranges from $0.01\ \text{cm}^2\text{V}^{-1}\text{s}^{-1}$ to $0.03\ \text{cm}^2\text{V}^{-1}\text{s}^{-1}$ ($V_{SD} = 15\ \text{V}$) showing that the difference is small and thus corroborating that the influence of the contacts is negligible.

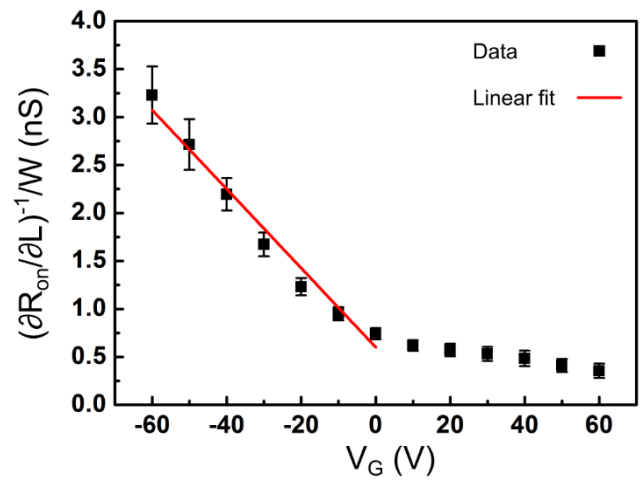

**Figure S3**: Contact resistance of metal/GNR interfaces. Width–normalized reciprocal slopes of the total resistance depending on gate voltage allowing for the determination of a contact resistance–corrected charge carrier mobility. Image adapted from Ref. 8.

## S5. Charge carrier density as a function of temperature

For the determination of the charge carrier density as

$$n_0 = \frac{eI\mu}{V} = \frac{e\mu}{R}, \tag{S3}$$

we combine I-V curves (**Fig. 3 (a)** in the main text) and transfer curves (**Fig. S4**) at variable temperatures. Here, we use the Ohmic part of the I-V curves to determine $R$ and we use $\mu_{FE}$ (measured at $V_D = 1$ V) as the mobility. At 260 K, the charge carrier density is high with approximately $2 \times 10^{12}$ cm$^{-2}$. This justifies the use of **Eq. 1** of the main text, which requires equal hopping steps through the device channel. Furthermore, the charge carrier density remains constant from 260 K down to approximately 100 K. Only at lower temperatures, the carrier density starts to decrease. Measuring the mobility at constant drain voltage is impeded at lower temperatures due to the large resistance.

The systematic error in the mobility, which we discuss above (**S3**), of course propagates to the charge carrier density. Nevertheless, the temperature dependence of the transfer curves is captured correctly in the field-effect mobility and therefore our results are robust against these systematic uncertainties.

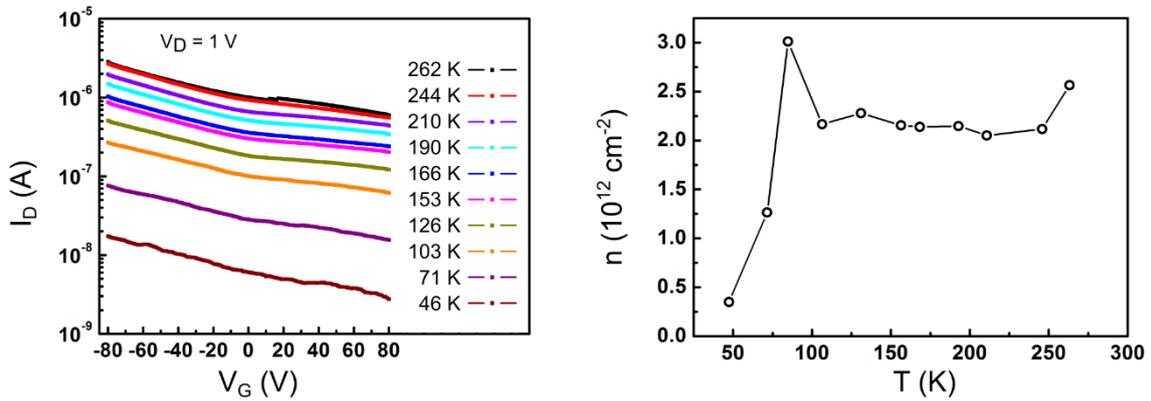

**Figure S4**. Temperature dependence of charge carrier density 5-AGNR network FETs. In (a), we present the temperature evolution of the transfer curves. The field-effect mobility is extracted from these curves and used to estimate the charge carrier density. As shown in (b), the charge carrier density is constant over a wide temperature range between 100 K and 260 K. Lines in are guides for the eye. Image adapted from Ref. 8.

## S6. Additional temperature dependent measurements

Additional to the charge transport data from the I-V curves shown in the main text, we performed a temperature sweep down to 5.6 K at fixed $V_D = 10$ V and $V_G = 0$ as shown in **Fig. S5**. These data are also included in the universal scaling shown in **Fig. 3 (c)** of the main text.

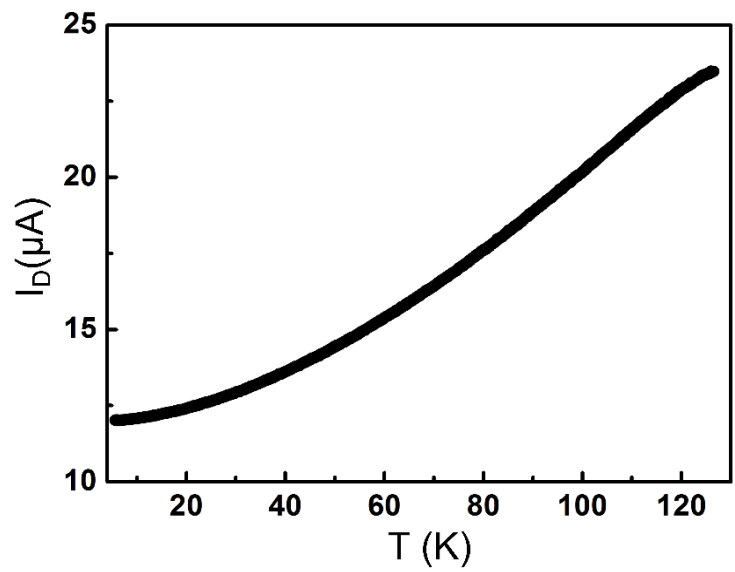

**Figure S5**. Temperature dependence of the channel current at $V_D = 10$ V and $V_G = 0$.